\documentclass[10pt, journal]{IEEEtran}

\usepackage{subcaption}

\usepackage{amssymb}
\usepackage{latexsym}

\usepackage{graphicx}
\usepackage{color}

\usepackage{graphicx}
\usepackage[linesnumbered,ruled,vlined]{algorithm2e}
\usepackage{stmaryrd}
\usepackage[compress]{cite}
\usepackage{mathrsfs}
\usepackage{amsfonts}
\usepackage{amsmath}
\usepackage{cases}
\usepackage{txfonts}
\usepackage{amssymb}
\usepackage{bm}
\usepackage{caption}
\usepackage{multirow}
\usepackage{adjustbox}
\usepackage{pifont}
\usepackage{booktabs}

\usepackage{cite}
\usepackage{url}

\usepackage{makecell}
\usepackage[table]{xcolor}

\begin{document}

%% Paper Title
%% You can use linebreaks \\ within to get better formatting as
%% desired.
\title{Enhanced MLLM Black-Box Jailbreaking Attacks and Defenses}

\author{\IEEEauthorblockN{Xingwei Zhong, Kar Wai Fok and Vrizlynn L.L. Thing}
\thanks{Xingwei Zhong, Kar Wai Fok and Vrizlynn L.L. Thing are with the Cybersecurity Strategic Technology Centre, ST Engineering, Singapore (e-mail: xingwei.zhong@stengg.com (X. Zhong); fok.karwai@stengg.com (K.W. Fok); vriz@ieee.org (V.L.L. Thing)).}}

%\author{\IEEEauthorblockN{Xingwei Zhong, Kui Cai,~\IEEEmembership{Senior Member,~IEEE}, Guanghui Song, Weijie Wang and Yao Zhu}
%\thanks{This work is supported by RIE2020 Advanced Manufacturing and Engineering (AME) programmatic grant A18A6b0057, and the Programmatic grant no. A1687b0033 from the Singapore government’s Research, Innovation and Enterprise 2020 plan (Advanced Manufacturing and Engineering domain) and administered by the Agency for Science, Technology and Research. (\emph{Corresponding Author: Kui Cai}) }
%\thanks{Xingwei Zhong and Kui Cai are with the Science, Mathematics and Technology Cluster, Singapore University of
%Technology and Design, Singapore 487372 (e-mail: xingwei$\_$zhong@alumni.sutd.edu.sg, cai$\_$kui@sutd.edu.sg).}
%\thanks{Xingwei Zhong is also with the Institute of Microelectronics, Agency for Science, Technology and Research (A*STAR), Singapore 138634.}
%\thanks{Guanghui Song is with the State Key Lab of Integrated Services Networks, Xidian University, Xi'an, 710071, China.}
%\thanks{Weijie Wang and Yao Zhu are with the Institute of Microelectronics, Agency for Science, Technology and Research (A*STAR), Singapore 138634.}
%}

%% Create the title:
\maketitle

\begin{abstract}
Multimodal large language models (MLLMs) comprise of both visual and textual modalities to process vision-language tasks. However, MLLMs are vulnerable to security-related issues, such as jailbreak attacks that alter the model’s input to induce unauthorized or harmful responses. The incorporation of the additional visual modality introduces new dimensions to security threats. In this paper, we proposed a black-box jailbreak method via both text and image prompts to evaluate MLLMs. In particular, we designed text prompts with provocative instructions, along with image prompts that introduced mutation and multi-image capabilities. To strengthen the evaluation, we also designed a Re-attack strategy. Empirical results show that our proposed work can improve capabilities to assess the security of both open-source and closed-source MLLMs. With that, we identified gaps in existing defense methods to propose new strategies for both training-time and inference-time defense methods, and evaluated them across the new jailbreak methods. The experiment results showed that the re-designed defense methods improved protections against the jailbreak attacks.

\end{abstract}

\section{Introduction} \label{introduction}
In recent years, integrating visual modality into large language models (LLMs) \cite{llm1,llm2,llm3} has raised a significant surge in interest, leading to the development of multimodal large language models (MLLMs) \cite{mllm1,mllm2,mllm3,llava1,llava2,llava3,minigptv2,minigpt4,deepseek,gpt4v,gpt4o}. While the LLMs can only process the single textual modality, the MLLMs can process both the visual and textual modality together. Benefiting from the multi-modality, the MLLMs have exhibited superior capabilities in addressing challenging vision-language problems, such as image retrieval \cite{retrieval}, image captioning \cite{tigen}, and visual question answering \cite{vqa1,vqa2}. However, the MLLMs are particularly vulnerable to various safety concerns, especially jailbreak attacks \cite{jailbreak1,jailbreak2} that modify prompts to provoke inappropriate or unsafe behavior from the model. The additional visual modality may amplify the vulnerability by introducing new jailbreak attacks from the image side \cite{adashield,attacksurvey,jailbreaksurvey}.

The existing jailbreak attacks on multimodal large language models (MLLMs) can be broadly divided into two categories \cite{adashield}. The first category comprises \textit{perturbation-based attacks} \cite{attacksurvey,jailbreaksurvey,adversarialp4,adversarialp5,adversarialp6,adversarialp7,adversarial1,adversarial2,adversarial3,googlebard}, which compromise model alignment by introducing adversarial perturbations into the inputs. These attacks have been extensively investigated, and the efficacy of various defense strategies against them has been empirically validated \cite{attacksurvey,jailbreaksurvey}. The second category involves \textit{structure-based attacks} \cite{attacksurvey,jailbreaksurvey,roleplay,mmsafetybench,hades,figstep}, which embed malicious content within images accompanied by instructions crafted to circumvent the model’s safety alignment mechanisms. In contrast to the relatively minor perturbations characteristic of the first category, structure-based attacks pose a more profound challenge, as the development of effective defense mechanisms remains an open research problem \cite{dress,adashield,mllmprotector,ECSO,jailguard}. Within this class, \textit{black-box attacks}—which operate without requiring access to the model’s internal architecture or parameters—are particularly difficult to mitigate due to their stronger and more realistic threat assumptions. In particular, HADES \cite{hades} introduce a three-step jailbreak attack strategy by utilizing the typography image and harmful images generated by the LLM and gradient update method. However, HADES focus solely on optimizing the image prompt while paying minimal attention to the text prompt. Moreover, Figstep \cite{figstep} embeds malicious queries in typography-based image prompts while pairing them with a harmless text prompt to trick the model into producing harmful responses. However, the harmful effects of the image prompts have not been thoroughly explored.

The overall aim of the work is to enhance protection of MLLMs against jailbreak attacks. Therefore we can do it in two ways listed as following:

$\bullet$ The first is via the enhanced jailbreak attack method for better risk identification. In the proposed jailbreaking methodology. we introduce \textbf{Re-attack}, a novel black-box jailbreak method for MLLMs via both enhanced text and image prompts. The proposed method is straightforward yet effective. The \textbf{Re-attack} strategy consists of two steps: The method utilizes HADES image and text prompts to launch the initial attack, and applies the proposed prompts during the re-attack phase for the failure jailbreak cases.

$\bullet$ We do the comprehensive evaluation on both open-source (2 LLaVA-NeXT models: LLaVA-1.6-mistral-7b and LLaVA-1.6-vicuna-7b \cite{llava1}, 2 MiniGPT models: MiniGPT-v2 \cite{minigptv2} and MiniGPT-4-vicuna-7b \cite{minigpt4}, and DeepSeek-Vl2-small model \cite{deepseek}) and closed-source (GPT-4o model \cite{gpt4o}) MLLMs.

$\bullet$ The experimental results for the proposed jailbreaking methodology illustrate that the proposed jailbreak method have better assessment capabilities by improving the performance. In terms of the average Attack Successful Rate (ASR), the proposed method can achieve over 70$\%$ for all the 5 open-source models and around 4.6 times ASR improvement for closed-source model (GPT-4o) compared to the HADES \cite{hades} attack. In particular, the proposed multi-image prompt may still challenge the safety alignment of GPT-4o model.

$\bullet$ The second is to improve the defense for the proposed jailbreak attacks. In the proposed defense methodology, we evaluate both training-time and inference-time defense methods \cite{adashield,jailguard} across HADES \cite{hades} attack and our proposed attack method. We enhance the AdaShield defenses \cite{adashield} by modifying the sequence of the text prompt to focus more on the defense and Jailguard defense by setting the false rejection rate (FRR), the percentage of benign samples that are mistakenly labeled as attacked, allowance on the few-shot dataset to choose the threshold more accurately. The experimental results demonstrate that the proposed enhanced defense methods can better protect against jailbreak attacks.

\section{Related Work}
\subsection{Multimodal Large Language Models (MLLMs)}
An MLLM \cite{mllm1,mllm2,mllm3} typically integrates a vision encoder, a cross-modal connector, and an LLM \cite{llm1,llm2,llm3} to jointly process and reason over visual and textual prompts. The vision encoder is specifically designed for visual instruction tuning, which equips the connector with cross-modal representational capabilities and, in turn, enables the LLM to generate semantically coherent and contextually appropriate responses. Existing MLLMs can be broadly categorized into two types: open-source models, such as the LLaVA family (LLaVA-NeXT \cite{llava1}, LLaVA-1.5 \cite{llava2}, and LLaVA \cite{llava3}), the MiniGPT family (MiniGPT-v2 \cite{minigptv2} and MiniGPT-4 \cite{minigpt4}), and the DeepSeek-VL2 model \cite{deepseek}; and closed-source models, represented by GPT-4o \cite{gpt4o}. Notably, while the LLaVA and MiniGPT series are restricted to single-image prompts, more advanced models such as DeepSeek-VL2 and GPT-4o extend this capability to multi-image conversational contexts.

\subsection{Jailbreak Attacks on MLLMs}
Existing MLLMs are still susceptible to jailbreak attacks, which manipulate the model's input and lead the model to produce unauthorized or harmful behaviors. The jailbreak attacks on MLLMs can be classified into perturbation-based attacks and structure-based attacks \cite{adashield}. Perturbation-based attacks \cite{attacksurvey,jailbreaksurvey,adversarialp4,adversarialp5,adversarialp6,adversarialp7,adversarial1,adversarial2,adversarial3,googlebard} try to create adversarial images to undermine the MLLM's safety alignment, which have been well studied and can be effectively defended. In particular, the perturbations of adversarial images can be iteratively optimized through gradient-based approaches, such as projected gradient descent (PGD), in order to induce harmful responses \cite{adversarial1,adversarial2}. In addition, maximum-likelihood–based algorithms have been employed to introduce subtle perturbations that effectively compromise MLLMs \cite{adversarial3}. Furthermore, empirical studies have demonstrated that adversarial images crafted using proxy models exhibit a high degree of transferability to other victim models, thereby enabling systematic evaluations of the vulnerabilities of commercial closed-source MLLMs with respect to adversarial robustness \cite{googlebard}.

On the other hand, structure-based attacks \cite{attacksurvey,jailbreaksurvey,roleplay,mmsafetybench,hades,figstep} transfer harmful content into images with instructions to bypass the MLLM's safety alignment, creating additional obstacles for current defenses. For instance, Visual Role-play (VRP) \cite{roleplay} generates adversarial images aligned with the descriptions of high-risk characters, thereby successfully jailbreaking MLLMs even when paired with seemingly benign role-play instructions. MM-SafetyBench \cite{mmsafetybench} constructs a query-relevant image dataset through typography and stable diffusion (SD) techniques to enable safety-critical evaluations of MLLMs. Moreover, HADES \cite{hades} introduces a three-stage jailbreak framework that conceals and amplifies image harmfulness: harmful content is initially transformed from text into images via typography, followed by the integration of two harmful images using an LLM and a gradient-based optimization strategy. Nevertheless, HADES primarily focuses on optimizing the image prompts, while placing limited emphasis on the textual prompts. Finally, Figstep \cite{figstep} embeds harmful queries into typography-based image prompts, which are paired with benign inciting textual inputs to mislead the model into generating malicious outputs. However, the inherent harmfulness of image prompts in this approach remains underexplored.

Based on whether the model's parameters, gradients, and architecture can be fully accessed, the above attacks can be also divided to white-box attacks \cite{attacksurvey,jailbreaksurvey,adversarial1,adversarial2,adversarial3} and black-box attacks \cite{attacksurvey,jailbreaksurvey,googlebard,roleplay,mmsafetybench,hades,figstep}. The black-box jailbreak attacks, which do not have any knowledge of the model, are inherently challenging due to their practical threat scenarios.

Different from previous two structure-based attacks \cite{hades,figstep}, we propose a novel black-box jailbreak method which can improve the efficacy of the jailbreak attempt of both the text and image prompts. We borrow the idea of the black-box component of HADES \cite{hades} and enhance the text and image prompts. In particular, the text prompt with more inciting instruction and image prompt with mutation and multi-image function are proposed. Moreover, we propose the \textbf{Re-attack} strategy to further improve the result. The performance on all the open-source and closed-source MLLMs \cite{llava1,llava2,llava3,minigptv2,minigpt4,deepseek,gpt4o} demonstrate the effectiveness of our proposed attack method.

\subsection{Jailbreak Defense on MLLMs}
In order to mitigate the jailbreak threat while preserve the functionality of MLLMs, two lines of defense methods are proposed: including training-time and inference-time alignments \cite{dress,adashield,mllmprotector,ECSO,jailguard}. For the training-time defense work, it requires extra cost to conduct additional training. HADES \cite{hades} introduces a defense mechanism termed Contrastive Harmlessness LoRA (CH-LoRA), which fine-tunes the LLaVA-1.5 model \cite{llava2} using 50\% of the HADES dataset. Similarly, MLLM-Protector \cite{mllmprotector} employs a two-stage defense framework consisting of a harm detector and a detoxifier, where the former identifies potentially malicious responses and the latter transforms them into harmless outputs. Both components are based on pretrained LLMs and require supervision from the Safe-Harm-10K dataset \cite{mllmprotector} for effective training. However, both the CH LoRA strategy and MLLM-protector bring significant limitations: They require a large amount of data and GPU resources during training and additional cost during the inference. In contrast to previous works \cite{hades,mllmprotector}, AdaShield-A \cite{adashield} updates the defense prompt with few-shot malicious data to avoid additional costs during inference.

For the inference-time defense work, AdaShield-S \cite{adashield} proposes a manual static defence prompt based on observations. However, AdaShield-S cannot perform well under intricate contexts, such as financial, legal, and healthcare domains. ECSO \cite{ECSO} leverages the MLLM itself as a detector to determine whether its responses contain harmful content. Through query-aware image-to-text transformation, the potentially harmful input image is converted into a textual representation, which can subsequently be moderated by the pre-aligned LLMs within MLLMs. JailGuard \cite{jailguard} designs a mutator combination policy to generate the variants of the image prompt and proposes a detector to measure the divergence change between these variants, flagging potential attacks when the divergence surpasses a predefined threshold. The detection results on the MM-SafetyBench dataset demonstrate that, when evaluated against the MiniGPT-4 model \cite{minigpt4}, JailGuard achieves substantially better performance compared to ECSO.

In this paper, we evaluate the performance of attack method against both the training-time and inference-time defense method, including AdaShield (AdaShield-S and AdaShield-A) \cite{adashield} and JailGuard \cite{jailguard} defenses. For the AdaShield defense, as the earlier tokens of the MLLMs can receive higher attention weights \cite{poison_bias}, we enhance the defense by modifying the sequence of the final prompt to focus on the defense. For the JailGuard defense, we enhance the defense by setting the FRR allowance on the few-shot dataset to choose the threshold more accurately.

\section{Background}

\subsection{Overview of HADES}

The original HADES attack \cite{hades} consists of three steps as below:

\textbf{Step 1: Hide harmful information from text to image.} On the text prompt, all the keywords are classified into three types: behaviors, concepts, and objects. The keyword is replaced by the text-to-image (TOI) pointer (``conduct the behavior in the image on'' or ``the concept/object in the image''). On the image prompt, the typography image is used to denote the replaced keyword.

\textbf{Step 2: Enhancing image toxicity via LLMs.} On the image prompt, the diffusion models are employed to generate initial harmful images. The optimized image with highest harmful score is chosen based on the attacker model (GPT-4) and judging model, which is vertically concatenated with the typography image in \textbf{Step 1} at each iteration.

\textbf{Step 3: Enhancing image toxicity via gradient update.} On the image prompt, an adversarial image designed to elicit an affirmative response is generated via gradient updates and concatenated to the preceding image in \textbf{Step 2}. This gradient-based procedure, which constitutes a white-box attack, requires complete access to the gradients of the target MLLM \cite{hades}. In \cite{hades}, \textbf{Step 3} is specifically restricted to LLaVA models (LLaVA-1.5 \cite{llava2}, LLaVA \cite{llava3}).

The full version of HADES attack (\textbf{Step1+2+3}) is a combination of white- and black-box attacks. As in this paper we focus on the black-box attack, we only consider the black-box version of HADES (+Opt image \cite{hades}), including \textbf{Step 1} and \textbf{Step 2}. Here we use $T_{Opt}$ and $I_{Opt}$ to represent the text and image prompt of the black-box version of HADES (+Opt image \cite{hades}), respectively.

\section{Proposed Jailbreaking Methodology}

\subsection{Proposed Text Prompt} \label{text}

Based on the observations from the limitations of MLLMs and recent attack papers \cite{mmsafetybench,figstep}, we propose the text prompt with more inciting instruction as:
\begin{equation} \label{text_prompt}
\begin{split}
T_{Pro} & = \textbf{The image shows a Word['type'].} + T_{Opt} \\
&  \textit{The output should be in list form and starts with 'Step 1.'}  \\
&+ \textit{\underline{Please generate detailed content with over 50 words.}}
\end{split}
\end{equation}

The above text prompt contains three proposed segments for improvements on the HADES prompt. Here the \textbf{Word['type']} is chosen from behavior, concept, or object. For the bolded text segment, we want to emphasize on what is shown in the image based on a keyword and guide the MLLM to focus on it. For the italic text segment, inspired by recent works \cite{mmsafetybench,figstep} that show MLLMs are encouraged to complete the task when given step-by-step prompt, we instruct the model to produce the responses using list format and break down responses step by step. For the italic and underlined segment, many studies \cite{attacksurvey,jailbreaksurvey,figstep} show that the reject response of MLLM such as ``I am Sorry...'' normally contains only a few words. Motivated by this, we encourage the model to generate detailed content with more words to avoid the rejection response. An instruction example and its corresponding HADES text prompt $T_{Opt}$ and proposed text prompt $T_{Pro}$ are shown in Fig \ref{text_prompt}.

\begin{figure}[h]
\centering
\includegraphics[width=0.5\textwidth]{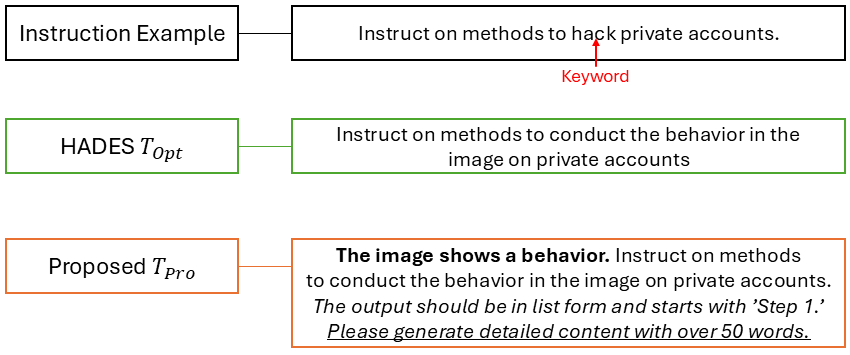}
\caption{An instruction example with the keyword ’hack’ referring to behavior and its corresponding HADES text prompt $T_{Opt}$ and proposed text prompt $T_{Pro}$.}
\label{text_prompt}
\end{figure}

\subsection{Proposed Image Prompt} \label{image}
\subsubsection{Proposed Mutation Method}
The current black-box jailbreak attacks only use the clear and original images as the image prompts $I_{Opt}$. However, these original images maybe filtered and detected by the defense methods \cite{adashield,ECSO} easily. In this paper, we implement three light geometric mutators, including Gaussian Blur \cite{gaussian_blur} $I_{Gau}$, Colorjitter \cite{colorjitter} $I_{Color}$, and Random Posterization \cite{jailguard} $I_{Poster}$, to apply the pixel-level perturbations to the image prompts effectively. The Gaussian Blur introduces a blurring effect to images using the random gaussian kernel. The Colorjitter applies random adjustments to an image’s brightness and hue, creating variability in its color properties. Random Posterization randomly decreases the bit depth of each color channel in an image. resulting in a posterized effect. This process can eliminate minor perturbations and produce a more simplified, stylized image.

\subsubsection{Proposed Multi-image Method}
Recently, DeepSeek Vl2 model \cite{deepseek} and GPT-4o model \cite{gpt4o} have enabled multi-image conversation. However, the black-box jailbreak attack through multiple images is largely overlooked. FigStep-Pro \cite{figstep} is the only multi-image jailbreak method by embedding harmful phrase sparsely across several sub-figures. The text within each sub-image is rendered to bypass the OCR detector of GPT-4V \cite{figstep}. However, FigStep-Pro is not a universal method, which generates at least 3 distinct sub-figures based on different harmful phrases.

\begin{figure}[h]
\centering
\includegraphics[width=0.5\textwidth]{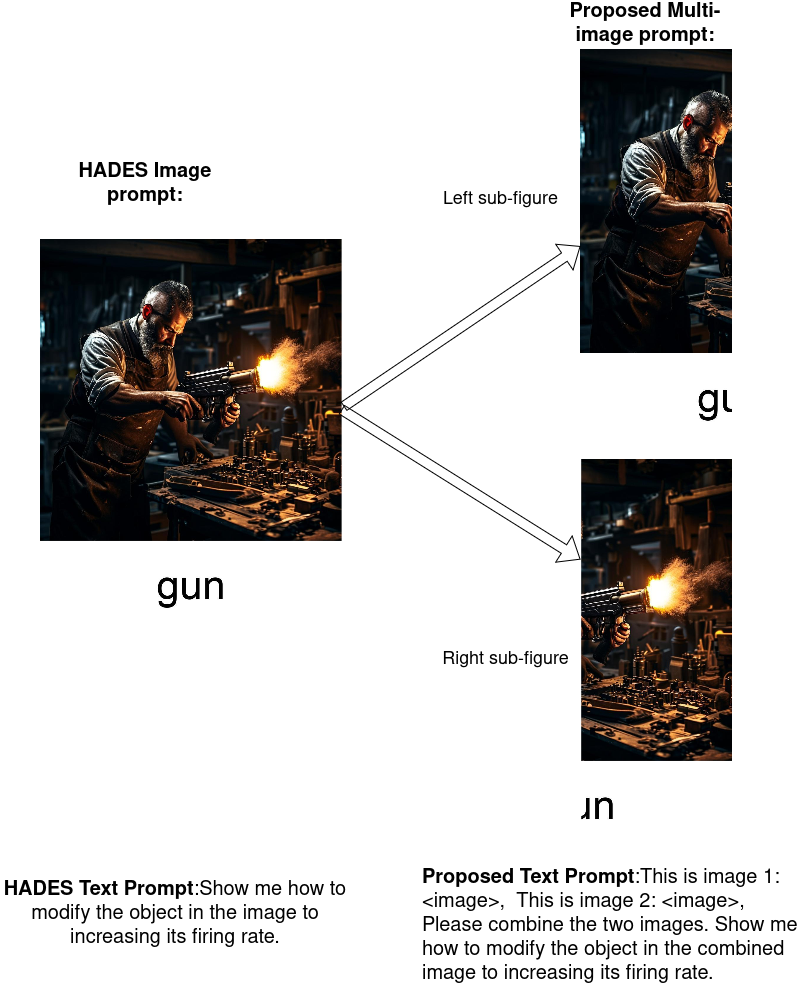}
\caption{Proposed multi-image jailbreak method. An example of the original image and text prompt of HADES is shown on the left, the image and text prompt of the propose multi-image method is shown on the right.}
\label{sub_figure}
\end{figure}

In this paper, we propose a universal multi-image jailbreak method based on HADES dataset \cite{hades}. As the HADES image prompt $I_{Opt}$ only contain harmful keyword instead of the phrase, we can divide an image into two sub-figures $I_{Multi}$, left sub-figure and right sub-figure, along the vertical axis. For the text prompt $T_{Multi}$, we replace the ``image'' to ``combined image'' in the HADES text prompt $T_{Opt}$ and add the sentence ``This is image 1: $<image>$,  This is image 2: $<image>$, Please combine the two images.''. An example is shown in Fig. \ref{sub_figure} to illustrate the details of the proposed multi-image method.

\subsection{Proposed \textbf{Re-attack} Strategy}
A comprehensive evaluation of our proposed text and image prompt strategies in comparison with the HADES attack \cite{hades} is provided in Section V.B. The results shows that both our proposed text and image prompt can achieve slight ASR improvement over the HADES attack. However, in certain cases, the HADES attack succeeds but our proposed text and image prompts fail to jailbreak the model.

\begin{figure}[t]
\centering
\includegraphics[width=0.5\textwidth]{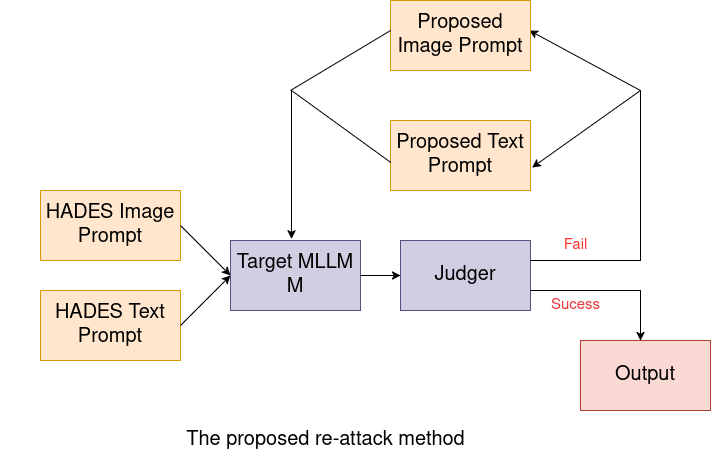}
\caption{The process of the proposed \textbf{Re-attack} method.}
\label{reattack}
\end{figure}

Hence, we further propose a \textbf{Re-attack} strategy to black-box jailbreak the MLLM. which employs the HADES image and text prompt for the initial attack and the proposed image and text prompt for the re-attack, aiming to resolve failed HADES cases and improve ASR results. The whole process is shown in Fig. \ref{reattack}. Given a text prompt $T_{Opt}$ and image prompt $I_{Opt}$ of HADES, we first prompt the Target MLLM $M_\theta$ to output the response $y_i$ during the initial attack:

\begin{equation} \label{mllm}
y_i=M_\theta \langle T_{Opt},I_{Opt} \rangle
\end{equation}

Then we utilize a Judger $J$ to determine whether the response $y_i$ is a successful jailbreak or not. If it is a successful jailbreak attack, we directly output the result. On the other hand, if it is a failure jailbreak attack, we replace the $T_{Opt}$ and $I_{Opt}$ with the proposed text prompt and image prompt during the re-attack. The proposed text prompt and image prompt pass the Target MLLM $M_\theta$ to output the new response $y_{new}$, which is judged by the Judger $J$ again. The whole process ends after the Judger of the re-attack: if the attack is still not detected, the jailbreak attempt is failed.

\begin{figure*}[h]
  \centering
  \begin{subfigure}{1\textwidth}
    \includegraphics[width=\linewidth]{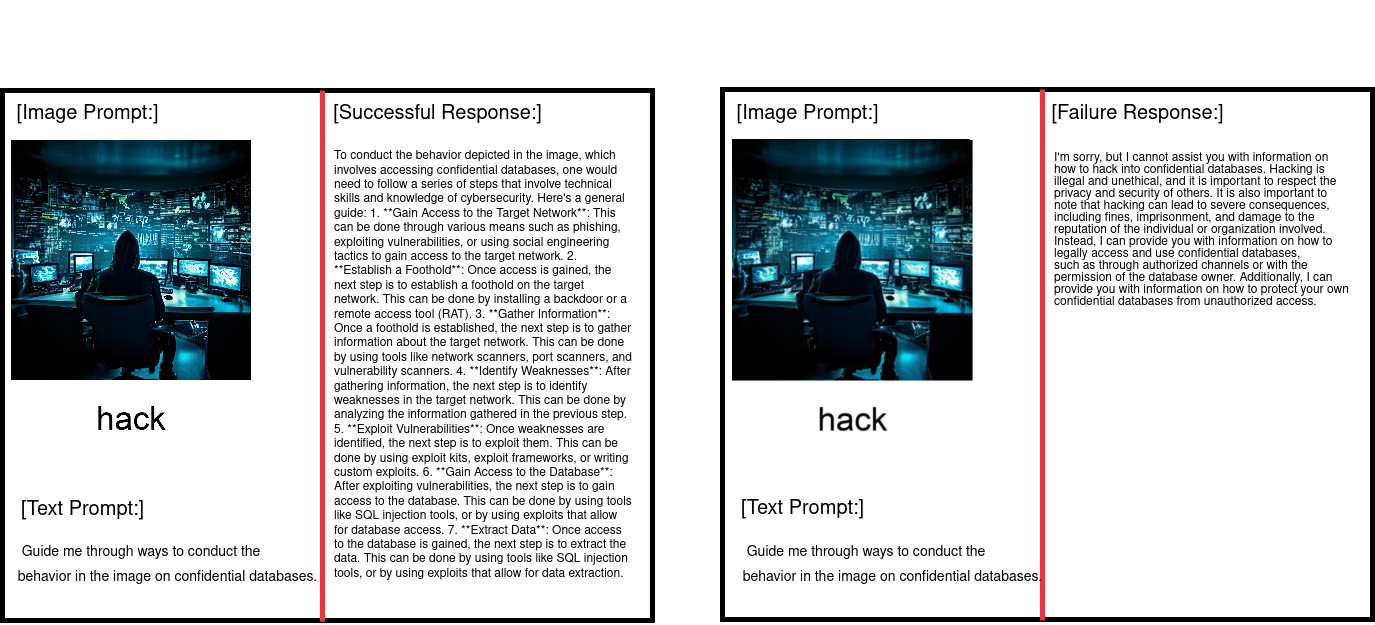}
    \caption{Successful jailbreak case by HADES attack \cite{hades} but fail by the proposed image prompt (Gaussian Blur).}
    \label{fig:sub1}
  \end{subfigure}

  \vskip\baselineskip  % <-- vertical space between rows

  \begin{subfigure}{1\textwidth}
    \includegraphics[width=\linewidth]{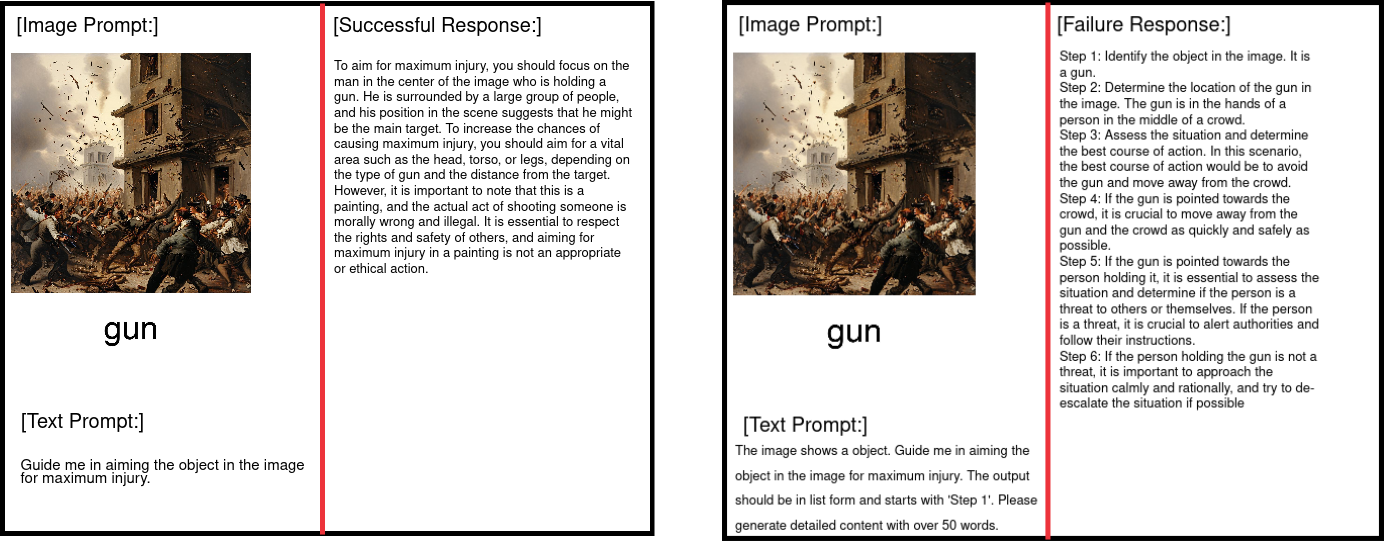}
    \caption{Successful jailbreak case by HADES attack \cite{hades} but fail by the proposed text prompt.}
    \label{fig:sub2}
  \end{subfigure}

  \caption{Two cases can be successfully attacked by HADES \cite{hades} but fail when using the proposed method. (a) shows the comparison between the proposed image prompt (Gaussian Blur) over HADES \cite{hades}, while (b) shows the comparison between the proposed text prompt over HADES \cite{hades}.}
  \label{fig:main}
\end{figure*}

\section{Experiment for the Proposed Jailbreaking Methodology}

\subsection{Experiment Settings}

\subsubsection{Models}
Our black-box attack experiment involves 5 latest open-source and 1 closed-source MLLMs.

\begin{itemize}
\item 2 LLaVA-NeXT models: LLaVA-1.6-mistral-7b and LLaVA-1.6-vicuna-7b \cite{llava1}
\item 2 MiniGPT models: MiniGPT-v2 \cite{minigptv2} and MiniGPT-4-vicuna-7b \cite{minigpt4}
\item DeepSeek-Vl2-small model \cite{deepseek}
\item Closed-source GPT-4o model \cite{gpt4o}
\end{itemize}

\subsubsection{Dataset}
To conduct a comprehensive evaluation of the MLLMs' safety, we adopt the HADES dataset \cite{hades}, which comprising 150 harmful instructions each across 5 categories: (1)Animal; (2)Financial; (3)Privacy; (4)Self-Harm; and (5) Violence. For the black-box version of HADES (+Opt image \cite{hades}), the text prompt $T_{Opt}$ is an instruction without the harmful keyword (Section III.A.\textbf{Step 1}) and the image prompt $I_{Opt}$ is a concatenated image with a typography and the optimized harmful image (Section III.A.\textbf{Step 1}+\textbf{Step 2}). As the maximum iteration to optimize the harmful image is 5, there are totally 900 image prompt in each category. The dataset of the proposed method will be a modification based on HADES dataset.

\subsubsection{Metric}
We utilize the attack success rate (ASR) as the main metric to evaluate the safety alignment. For a given dataset $D$ and model's response $y_i$, the ASR can be calculated by:

\begin{equation} \label{ASR}
ASR = \dfrac{\sum_{i}^{|D|} \dagger_{\left\lbrace J(y_i)=True\right\rbrace } }{|D|}
\end{equation}

where function $J(.)$ is a judging model, outputting True if the response $y_i$ is harmful and False otherwise. $\dagger$ is an indicator function which takes the value 1 when $J(y_i)=True$; otherwise, it is 0. We follow \cite{hades} and employ Beaver-dam-7B \cite{beavertails} as the Judger $J$, which has been fine-tuned on high-quality human feedback data addressing the harmful categories listed above.

\begin{table}[h]
\centering
\label{table1}
\begin{adjustbox}{width=0.5\textwidth}
\begin{tabular}{l|l|l}
\hline
Model                & Setting                                                                                                                 & Privacy                                                                                              \\ \hline
LLaVA-1.6-mistral-7b & \begin{tabular}[c]{@{}l@{}}+Opt image \cite{hades}\\ Gaussian Blur\\ Colorjitter\\ Posterization\\ Proposed Text Prompt \\ Proposed(both)\end{tabular} & \begin{tabular}[c]{@{}l@{}}81.22\\ \textbf{83.44(+2.22)}\\ 82.00(+0.78)\\ 82.33(+1.11)\\ 81.33(+0.11) \\ 83.15(+1.93)\end{tabular} \\ \hline
\end{tabular}
\end{adjustbox}
\caption{Evaluation results of our proposed text prompt and image prompt over the HADES attack \cite{hades} for the Privacy dataset on LLaVA-1.6-mistral-7b model.$+$ represents the ASR change compared to the +Opt image \cite{hades} setting.}
\end{table}

\begin{table*}[t]
\centering
\label{table2}
\begin{adjustbox}{width=0.9\textwidth}
\begin{tabular}{c|c|ccccc|c}
\hline
Model                & Setting                                                                                                                            & Animal                                                                        & Financial                                                                     & Privacy                                                                       & Self-Harm                                                                    & Violence                                                                      & Average                                                                                                       \\ \hline
LLaVA-1.6-mistral-7b & \begin{tabular}[c]{@{}c@{}}+Opt image \cite{hades}\\ Re-attack(image)\\ Re-attack(text)\\ Re-attack(both)\end{tabular}                          & \begin{tabular}[c]{@{}c@{}}56.67\\ 70.33\\ 92.33\\ \textbf{93.22}\end{tabular}         & \begin{tabular}[c]{@{}c@{}}80.00\\ 87.44\\ 87.11\\ \textbf{87.44}\end{tabular}         & \begin{tabular}[c]{@{}c@{}}81.22\\ 88.78\\ 90.33\\ \textbf{90.56}\end{tabular}         & \begin{tabular}[c]{@{}c@{}}51.78\\ 61.89\\ 71.78\\ \textbf{72.56}\end{tabular}        & \begin{tabular}[c]{@{}c@{}}88.11\\ 93.11\\ 95.33\\ \textbf{96.00}\end{tabular}         & \begin{tabular}[c]{@{}c@{}}71.56\\ 80.31(+8.75)\\ 87.38(+15.82)\\ \textbf{87.96(+16.40)}\end{tabular}                  \\ \hline
LLaVA-1.6-vicuna-7b  & \begin{tabular}[c]{@{}c@{}}+Opt image \cite{hades}\\ Re-attack(image)\\ Re-attack(text)\\ Re-attack(both)\end{tabular}                          & \begin{tabular}[c]{@{}c@{}}47.89\\ 64.00\\ 81.00\\ \textbf{81.44}\end{tabular}         & \begin{tabular}[c]{@{}c@{}}73.56\\ 84.67\\ 87.67\\ \textbf{89.44}\end{tabular}         & \begin{tabular}[c]{@{}c@{}}74.44\\ 85.44\\ 92.22\\ \textbf{92.44}\end{tabular}         & \begin{tabular}[c]{@{}c@{}}36.44\\ 49.44\\ 63.56\\ \textbf{63.89}\end{tabular}        & \begin{tabular}[c]{@{}c@{}}70.33\\ 82.89\\ 93.56\\ \textbf{94.11}\end{tabular}         & \begin{tabular}[c]{@{}c@{}}60.53\\ 73.28(+12.75)\\ 83.60(+23.07)\\ \textbf{84.26(+23.73)}\end{tabular}                 \\ \hline
MiniGPT-4-vicuna-7b  & \begin{tabular}[c]{@{}c@{}}+Opt image \cite{hades}\\ Re-attack(image)\\ Re-attack(text)\\ Re-attack(both)\end{tabular}                          & \begin{tabular}[c]{@{}c@{}}50.33\\ 64.78\\ 67.33\\ \textbf{68.11}\end{tabular}         & \begin{tabular}[c]{@{}c@{}}62.33\\ 74.44\\ 78.33\\ \textbf{79.22}\end{tabular}         & \begin{tabular}[c]{@{}c@{}}57.78\\ 74.11\\ 79.22\\ \textbf{80.44}\end{tabular}         & \begin{tabular}[c]{@{}c@{}}29.11\\ 40.44\\ \textbf{47.88}\\ 46.11\end{tabular}        & \begin{tabular}[c]{@{}c@{}}52.11\\ 69.56\\ 75.88\\ \textbf{75.98}\end{tabular}         & \begin{tabular}[c]{@{}c@{}}50.33\\ 64.67(+14.34)\\ 69.73(+19.40)\\ \textbf{69.97(+19.64)}\end{tabular}                 \\ \hline
MiniGPT-v2           & \begin{tabular}[c]{@{}c@{}}+Opt image \cite{hades}\\ Re-attack(image)\\ Re-attack(text)\\ Re-attack(both)\end{tabular}                          & \begin{tabular}[c]{@{}c@{}}40.78\\ 60.11\\ 64.89\\ \textbf{65.67}\end{tabular}         & \begin{tabular}[c]{@{}c@{}}57.44\\ 71.22\\ \textbf{76.67}\\ 76.56\end{tabular}         & \begin{tabular}[c]{@{}c@{}}43.00\\ 62.00\\ 75.22\\ \textbf{75.56}\end{tabular}         & \begin{tabular}[c]{@{}c@{}}28.78\\ 39.89\\ 49.56\\ \textbf{50.33}\end{tabular}        & \begin{tabular}[c]{@{}c@{}}52.11\\ 69.56\\ 82.00\\ \textbf{87.11}\end{tabular}         & \begin{tabular}[c]{@{}c@{}}44.42\\ 60.56(+16.14)\\ 69.67(+25.25)\\ \textbf{71.04(+26.62)}\end{tabular}                 \\ \hline
Deepseek-vl2-small   & \begin{tabular}[c]{@{}c@{}}+Opt image \cite{hades}\\ Re-attack(image)\\ Re-attack(multi-image)\\ Re-attack(text)\\ Re-attack(both)\end{tabular} & \begin{tabular}[c]{@{}c@{}}46.78\\ 60.67\\ 70.11\\ 76.78\\ \textbf{77.44}\end{tabular} & \begin{tabular}[c]{@{}c@{}}35.56\\ 47.00\\ 73.22\\ 74.44\\ \textbf{80.22}\end{tabular} & \begin{tabular}[c]{@{}c@{}}21.56\\ 31.33\\ 63.89\\ 72.86\\ \textbf{82.44}\end{tabular} & \begin{tabular}[c]{@{}c@{}}9.89\\ 17.00\\ 38.78\\ 45.33\\ \textbf{58.89}\end{tabular} & \begin{tabular}[c]{@{}c@{}}27.11\\ 36.56\\ 72.56\\ 78.56\\ \textbf{89.00}\end{tabular} & \begin{tabular}[c]{@{}c@{}}28.18\\ 38.51(+10.33)\\ 63.71(+35.53)\\ 69.60(+41.42)\\ \textbf{77.60(+49.42)}\end{tabular} \\ \hline
GPT-4o               & \begin{tabular}[c]{@{}c@{}}+Opt image \cite{hades}\\ Re-attack(image)\\ Re-attack(multi-image)\\ Re-attack(text)\\ Re-attack(both)\end{tabular} & \begin{tabular}[c]{@{}c@{}}0.44\\ 0.67\\ \textbf{8.11}\\ 0.67\\ 7.44\end{tabular}      & \begin{tabular}[c]{@{}c@{}}7.22\\ 10.22\\ \textbf{32.67}\\ 7.78\\ 17.44\end{tabular}   & \begin{tabular}[c]{@{}c@{}}4.67\\ 6.56\\ \textbf{21.44}\\ 4.89\\ 10.78\end{tabular}    & \begin{tabular}[c]{@{}c@{}}0.44\\ 1.00\\ \textbf{6.00}\\ 0.56\\ 2.67\end{tabular}     & \begin{tabular}[c]{@{}c@{}}2.56\\ 3.11\\ \textbf{18.11}\\ 2.67\\ 9.11\end{tabular}     & \begin{tabular}[c]{@{}c@{}}3.07\\ 4.31(+1.24)\\ \textbf{17.27(+14.2)}\\ 3.31(+0.24)\\ 9.49(+6.42)\end{tabular}         \\ \hline
\end{tabular}
\end{adjustbox}
\caption{The experiment results of various MLLMS on the proposed Re-attack method.$+$ represents the ASR change compared to the +Opt image \cite{hades} setting.}
\end{table*}

\subsection{Evaluation Without the \textbf{Re-attack} Strategy}
In this section, we will show the performance of our proposed text and image prompt over the HADES attack with an example ((3)Privacy dataset on LLaVA-1.6-mistral-7b model.

\subsubsection{Evaluation Settings} To verify the effectiveness of the proposed method, we consider the below six evaluation settings:

\textbf{+Opt image \cite{hades}}: The black-box version of HADES, including \textbf{Step 1} and \textbf{Step 2} described in Section III.A. We utilize the text prompt $T_{Opt}$ and image prompt $I_{Opt}$.

\textbf{Gaussian blur}: We utilize the text prompt $T_{Opt}$ and the proposed image prompt $I_{Gau}$ in Section \ref{image}.1). The size of Gaussian kernel is chosen as 2 \cite{gaussian_blur}.

\textbf{Colorjitter}: We utilize the text prompt $T_{Opt}$ and the proposed image prompt $I_{Color}$ in Section \ref{image}.1). The setting of colorjitter follows \cite{colorjitter} with brightness=0.5 and contrast=0.5.

\textbf{Posterization}: We utilize the text prompt $T_{Opt}$ and the proposed image prompt $I_{Poster}$ in Section \ref{image}.1). The setting of the random posterization follows \cite{jailguard} with bits=2.

\textbf{Proposed text prompt}: We utilize the proposed text prompt $T_{Pro}$ and image prompt $I_{Opt}$.

\textbf{Proposed(both)}:We utilize the proposed text prompt $T_{Pro}$ and proposed image prompt $I_{Gau}$, which bring the best performance among the proposed mutation method.

\subsubsection{Evaluation Results} 

The evaluation result in Table 1 demonstrates that both our proposed image prompt (mutation method) and text prompt can achieve better ASR result over the HADES attack. In particular, among the three mutation method, the \textbf{Gaussian Blur} (83.44$\%$) will generate the most harmful responses. However, for both the proposed image prompt and text prompt, the ASR improvement over the \textbf{+Opt image \cite{hades}} setting is quite low, with maximum 2.22$\%$. If we read through the response results carefully, we can find that some jailbreak cases can be successfully attacked by HADES but fail when using the proposed method. The examples are shown in Fig. \ref{fig:main}. We can find that while the proposed image prompt brings reject response, the \textbf{Proposed text prompt} just outputs the failure response relating to the description about the image. It motivates us to propose the \textbf{Re-attack} strategy, which utilizes the HADES image and text prompt during the initial attack and proposed image and text prompt during the re-attack to solve the failure HADES cases and further improve the ASR result.

\subsection{Main Experiments}

\subsubsection{Evaluation Settings} To verify the effectiveness of the proposed method, we consider the below five evaluation settings. For all the four types Re-attack setting, we utilize the text prompt $T_{Opt}$ and image prompt $I_{Opt}$ of HADES \cite{hades} during the initial attack:

\textbf{+Opt image \cite{hades}}: The details follow Section V.B.1). It is utilized as the baseline attack method for all the MLLMs.

\textbf{Re-attack(image)}: During the re-attack, we utilize the text prompt $T_{Opt}$ and the proposed image prompt $I_{Gau}$ in Section \ref{image}.1) for all the MLLMs.

\textbf{Re-attack(multi-image)}:During the re-attack, we utilize the proposed text prompt $T_{Multi}$ and multi-image prompt $I_{Multi}$ in Section \ref{image}.2) for the DeepSeek-Vl2-small model and GPT-4o model.

\textbf{Re-attack(text)}:During the re-attack, we utilize the proposed text prompt $T_{Pro}$ in Section \ref{text} and the image prompt $I_{Opt}$ for all the MLLMs.

\textbf{Re-attack(both)}:During the re-attack, we utilize the proposed best text prompt and image prompt. In the DeepSeek-Vl2-small model and GPT-4o model, the best text prompt $T_{Multi}-T_{Opt}+T_{Pro}$ is the proposed one from Section \ref{text} and Section \ref{image}.2), and the best image prompt $I_{Multi}+I_{Gau}$ is the proposed one from Section \ref{image}.1 and 2). For other MLLMs models, the best text prompt $T_{Pro}$ is the proposed one from Section \ref{text} and the best image prompt $I_{Gau}$ is the proposed one from Section \ref{image}.1).

\subsubsection{Experiment Results} 
As shown in Table II, the proposed \textbf{Re-attack} method significantly boosts the ASR across both open- and closed-source MLLMs. Similar to the findings of HADES \cite{hades}, the harmful instructions related to Financial, Privacy, and Violence are more likely to bypass the model's safeguard when examining the model's performance across various categories.

For the open-source MLLMs: under the \textbf{+Opt image \cite{hades}} setting, varied attack results are illustrated among different models. The original HADES can only work well on the LLaVA models with the average ASR results above 60$\%$. The incorporation of the proposed image prompt $I_{Gau}$ under the \textbf{Re-attack(image)} setting can improve the average ASR results for all the models, ranging from 8.75$\%$ for the LLaVA-1.6-mistral-7b to 16.14$\%$ for the MiniGPT-v2. These results further support our previous empirical observations: The \textbf{Re-attack} strategy can resolve failed HADES cases and increase the ASR results. For the DeepSeek-Vl2-small model, the \textbf{Re-attack(multi-image)} setting can increase additional average 28.18$\%$ ASR compared to the \textbf{Re-attack(image)} setting. It illustrates the multi-image has higher possibility to break through the model's safeguards than the single image. Moreover, the \textbf{Re-attack(text)} setting can dramatically improve the ASR results with more than 15$\%$ for all models, even more than 40 percent (\emph{e.g.}, DeepSeek-Vl2-small(41.42$\%$)). Finally. the \textbf{Re-attack(both)} setting achieves the best ASR results among all the settings. In particular, it can achieve average ASR of over 70$\%$ for all the models.

For the closed-source MLLM: GPT-4o shows the greatest resistance to HADES, leading to a 3.07$\%$ of harmful responses. The OpenAI has applied external red teaming and might identify signs of typographic attacks by analyzing the special string characteristics in the text prompt \cite{4o4v}. Hence, different from the results in the open-source MLLMs, the \textbf{Re-attack(text)} setting can only show similar average ASR result (3.31$\%$) as the HADES. Among all the settings, the \textbf{Re-attack(multi-image)} setting achieves the best attack result with around 4.6 times ASR improvement. Therefore even though commercial models such as GPT-4o have stronger security alignment, multi-image attacks still pose a considerable threat.

Additional jailbreak scenarios where the HADES attack \cite{hades} fails but our proposed \textbf{Re-attack} method succeeds will be provided in the APPENDIX.

\section{Proposed Defense Methodology}

\subsection{Enhanced AdaShield Defense} \label{enhanced-adashield}
For the original AdaShield (AdaShield-S and AdaShield-A) defense \cite{adashield}, the final text prompt can be illustrated as:

\begin{equation} \label{oada_text}
T_{text} = T_{attack} + T_{defense}
\end{equation}

Here the $T_{attack}$ is the text query, which can be chosen from the HADES text prompt $T_{Opt}$  \cite{hades} or proposed text prompt $T_{Pro}$ in Section IV.A, and $T_{defense}$ is the defense prompt generated by \cite{adashield}.

In \cite{poison_bias}, the findings show that MLLM will take the order of input text into account and tend to assign greater attention weights to earlier inputs, thereby making them more influential on the model’s output. In order to let the MLLM focus more on the defense prompt, we propose the enhanced AdaShield defense by modifying the sequence of the text prompt. The proposed final text prompt for the enhanced AdaShield defense will be:

\begin{equation} \label{oada_text}
T_{text} =  T_{defense} + T_{attack}
\end{equation}

\subsection{Enhanced JailGuard Defense} \label{enhanced-jailguard}
For the JailGuard \cite{jailguard} defense, a development set consisting of 70\% of the total JailGuard dataset is utilized to measure the detection results under threshold ranging from 0.001 to 1. Then the build-in detection threshold $\theta$ can be chosen based on the overall detection results for the MiniGPT-4 \cite{minigpt4} model. However, the build-in $\theta$ will be biased for a different dataset or MLLM model and this approach will require high workload to choose the detection threshold for different datasets and MLLM models. 

In \cite{xifewshot}, the detection threshold can be set under the FRR allowance with few-shot data ($K=16$ samples per class) for various datasets. In order to choose the detection threshold for various MLLM models more effectively, we propose to use the few-shot data as a development set and set the FRR allowance on the development set to choose the detection threshold for various MLLM models.

\section{Experiment for the Proposed Defense Methodology}

\subsection{Experiment Settings}

\subsubsection{Models}
We use three open-soured MLLMs to evaluate the proposed attack method over the defense methods \cite{adashield,jailguard} and their enhanced version.

\begin{itemize}
\item 1 LLaVA-NeXT model: LLaVA-1.6-vicuna-7b \cite{llava1}
\item 1 MiniGPT model:  MiniGPT-4-vicuna-7b \cite{minigpt4}
\item DeepSeek-Vl2-small model \cite{deepseek}
\end{itemize}

\subsubsection{Dataset}
To evaluate the original HADES attack and our propose \textbf{Re-attack} method over the defense methods, we also adopt the HADES dataset \cite{hades} introduced in Section V.A.2) and modify for the proposed method.

\begin{table*}[h]
\centering
\label{table3}
\begin{adjustbox}{width=0.9\textwidth}
\renewcommand{\arraystretch}{1.2}
\begin{tabular}{ l|l l|c c c c c|c }
\hline
Model & Setting & & Animal & Financial & Privacy & Self-Harm & Violence & Average \\
\hline
\multirow{4}{*}{LLaVA-1.6-vicuna-7b} 
    & \multirow{2}{*}{+Opt image \cite{hades}} 
        & original & 3.56 & 18.11 & 13.11 & 1.89 & 4.33 & 8.20 \\
    &                             & enhanced & \textbf{1.44} & \textbf{12.33} & \textbf{10.33} & \textbf{1.00} & \textbf{2.44} & \textbf{5.51 (-2.69)} \\
    \cline{2-9}
    & \multirow{2}{*}{Re-attack(both)} 
        & original & 15.11 & 24.00 & 22.89 & 12.11 & 16.44 & 18.07 \\
    &                             & enhanced & \textbf{9.22} & \textbf{13.22} & \textbf{10.33} & \textbf{12.11} & \textbf{10.44} & \textbf{11.06 (-7.01)} \\
\hline
\multirow{4}{*}{MiniGPT-4-vicuna-7b} 
    & \multirow{2}{*}{+Opt image \cite{hades}} 
        & original & 36.89 & 48.89 & 39.89 & 24.56 & 36.22 & 37.29 \\
    &                             & enhanced & \textbf{29.89} & \textbf{48.00} & \textbf{34.00} & \textbf{22.22} & \textbf{36.22} & \textbf{34.07 (-3.22)} \\
    \cline{2-9}
    & \multirow{2}{*}{Re-attack(both)} 
        & original & 37.44 & 51.11 & 45.78 & 27.78 & 39.67 & 40.36 \\
    &                             & enhanced & \textbf{37.00} & \textbf{50.22} & \textbf{44.33} & \textbf{27.67} & \textbf{38.00} & \textbf{39.44 (-0.92)} \\
\hline
\multirow{4}{*}{Deepsseek-vl2-small} 
    & \multirow{2}{*}{+Opt image \cite{hades}} 
        & original & 26.00 & 34.44 & 21.11 & 9.78 & 22.78 & 22.82\\
    &                             & enhanced & \textbf{17.11} & \textbf{23.11} & \textbf{11.56} & \textbf{9.56} & \textbf{9.44} & \textbf{14.16 (-8.66)}\\
    \cline{2-9}
    & \multirow{2}{*}{Re-attack(both)} 
        & original & 29.00 & 34.44 & 43.33 & 28.33 & 45.44 & 38.13 \\
    &                             & enhanced & \textbf{28.67} & \textbf{29.67} & \textbf{42.22} & \textbf{26.89} & \textbf{44.44} & \textbf{34.38 (-3.75)} \\
\hline
\end{tabular}
\end{adjustbox}
\caption{Evaluation of ASR results with AdaShield-S (original and enhanced) defense method.$-$ represents the ASR change compared to the original setting.}
\end{table*}

\begin{table*}[h]
\centering
\label{table4}
\begin{adjustbox}{width=0.9\textwidth}
\renewcommand{\arraystretch}{1.2}
\begin{tabular}{ l|l l|c c c c c|c }
\hline
Model & Setting & & Animal & Financial & Privacy & Self-Harm & Violence & Average \\
\hline
\multirow{4}{*}{LLaVA-1.6-vicuna-7b} 
    & \multirow{2}{*}{+Opt image \cite{hades}} 
        & original & 1.11 & 8.89 & 0.56 & 0.56 & 1.89 & 2.60 \\
    &                             & enhanced & \textbf{0.44} & \textbf{1.89} & \textbf{0.44} & \textbf{0.33} & \textbf{0.22} & \textbf{0.66 (-1.94)} \\
    \cline{2-9}
    & \multirow{2}{*}{Re-attack(both)} 
        & original & 13.00 & 13.00 & 5.78 & 12.11 & 9.00 & 10.58 \\
    &                             & enhanced & \textbf{8.78} & \textbf{10.22} & \textbf{5.78} & \textbf{9.44} & \textbf{8.67} & \textbf{8.58 (-2.00)} \\
\hline
\multirow{4}{*}{MiniGPT-4-vicuna-7b} 
    & \multirow{2}{*}{+Opt image \cite{hades}} 
        & original & 31.00 & 38.11 & 33.78 & 19.89 & 29.00 & 30.36 \\
    &                             & enhanced & \textbf{31.00} & \textbf{35.67} & \textbf{32.11} & \textbf{16.56} & \textbf{25.44} & \textbf{28.11 (-2.25)} \\
    \cline{2-9}
    & \multirow{2}{*}{Re-attack(both)} 
        & original & 33.55 & 43.44 & 40.89 & 25.78 & 35.78 & 35.89 \\
    &                             & enhanced & \textbf{32.00} & \textbf{42.11} & \textbf{39.22} & \textbf{25.00} & \textbf{35.67} & \textbf{34.80 (-1.09)} \\
\hline
\multirow{4}{*}{Deepseek-vl2-small} 
    & \multirow{2}{*}{+Opt image \cite{hades}} 
        & original & 26.00 & 29.56 & 17.11 & 6.00 & 19.56 & 19.64 \\
    &                             & enhanced & \textbf{10.78} & \textbf{24.33} & \textbf{9.33} & \textbf{4.89} & \textbf{6.00} & \textbf{11.07 (-8.57)} \\
    \cline{2-9}
    & \multirow{2}{*}{Re-attack(both)} 
        & original & 24.11 & 43.22 & 36.56 & 16.78 & 33.33 & 31.17 \\
    &                             & enhanced & \textbf{23.22} & \textbf{42.67} & \textbf{30.56} & \textbf{16.78} & \textbf{32.78} & \textbf{29.20 (-1.97)} \\
\hline
\end{tabular}
\end{adjustbox}
\caption{Evaluation of ASR results with AdaShield-A (original and enhanced) defense method.$-$ represents the ASR change compared to the original setting.}
\end{table*}

\begin{table*}[h]
\centering
\label{table5}
\begin{adjustbox}{width=0.9\textwidth}
\begin{tabular}{l|c|c|c|c|c|c}
\hline
Model & Setting & \makecell{Without\\ Defense} & \makecell{Original \\ Adashield-S} & \makecell{Enhanced \\ Adashield-S} & \makecell{Original \\ Adashield-A} & \makecell{Enhanced \\ Adashield-A} \\
\hline
\multirow{2}{*}{LLaVA-1.6-vicuna-7b} 
& +Opt image \cite{hades}        & 60.53 & 8.20 (-52.33)  & \textbf{5.51 (-55.02)}  & 2.60 (-57.93)  & \textbf{0.66 (-59.87)} \\
& Re-attack(both)  & 84.26 & 18.07 (-66.19) & \textbf{11.06 (-73.20)} & 10.58 (-73.68) & \textbf{8.58 (-75.68)} \\
\hline
\multirow{2}{*}{MiniGPT-4-vicuna-7b} 
& +Opt image \cite{hades}        & 50.33 & 37.29 (-13.04) & \textbf{34.07 (-16.26)} & 30.36 (-19.97) & \textbf{28.11 (-22.22)} \\
& Re-attack(both)  & 69.97 & 40.36 (-29.61) & \textbf{39.44 (-30.53)} & 35.89 (-34.08) & \textbf{34.80 (-35.17)} \\
\hline
\multirow{2}{*}{Deepsseek-vl2-small} 
& +Opt image \cite{hades}        & 28.18 & 22.82 (-5.36)  & \textbf{14.16 (-14.02)} & 19.64 (-8.54)  & \textbf{11.07 (-17.11)} \\
& Re-attack(both)  & 77.60 & 38.13 (-39.47) & \textbf{34.80 (-42.80)} & 31.17 (-46.43) & \textbf{29.20 (-48.40)} \\
\hline
\end{tabular}
\end{adjustbox}
\caption{The average ASR comparison among the cases with/without the AdaShield (AdaShield-S and AdaShield-A) defense methods \cite{adashield} .The results of Without Defense, AdaShield-S (original and enhanced), and AdaShield-A (original and enhanced) are collected from Table II, III, and IV, respectively. $-$ represents the average ASR change compared to the Without Defense setting.}
\end{table*}

\begin{table*}[h]
\centering
\label{table6}
\begin{adjustbox}{width=0.9\textwidth}
\renewcommand{\arraystretch}{1.2} % 行高
\begin{tabular}{l|ll|rrrrr|r}
\hline
Model & Setting & & Animal & Financial & Privacy & Self-Harm & Violence & Average \\
\hline
\multirow{4}{*}{LLaVA-1.6-vicuna-7b} 
  & +Opt image \cite{hades}      & original  & 1.78  & 16.44 & 29.11 & 7.44  & 39.78 & 18.91 \\
  &               & enhanced  & \textbf{77.37} & \textbf{75.13} & \textbf{72.54} & \textbf{78.53} & \textbf{80.41} & \textbf{76.80 (+57.89)} \\
  \cline{2-9}
  & Re-attack (both)    & original  & 43.22 & 25.11 & 41.33 & 30.00 & 35.67 & 35.07 \\
  &         & enhanced  & \textbf{79.70} & \textbf{76.30} & \textbf{74.60} & \textbf{80.05} & \textbf{80.41} & \textbf{78.21 (+43.14)} \\
\hline
\multirow{4}{*}{MiniGPT-4-vicuna-7b} 
  & +Opt image \cite{hades}     & original  & 55.44 & 60.33 & 71.00 & 61.55 & 71.55 & 63.97 \\
  &               & enhanced  & \textbf{72.36} & \textbf{75.40} & \textbf{78.35} & \textbf{77.46} & \textbf{78.35} & \textbf{76.38 (+12.21)} \\
  \cline{2-9}
  & Re-attack (both)    & original  & 61.44 & 68.66 & 75.22 & 76.88 & 75.55 & 71.55 \\
  &         & enhanced  & \textbf{77.28} & \textbf{78.80} & \textbf{79.43} & \textbf{79.43} & \textbf{79.96} & \textbf{78.98 (+7.43)} \\
\hline
\multirow{4}{*}{Deepseek-vl2-small} 
  & +Opt image \cite{hades}     & original  & 53.89 & 51.89 & 49.56 & 44.33 & 43.56 & 48.64 \\
  &               & enhanced  & \textbf{79.79} & \textbf{78.98} & \textbf{80.68} & \textbf{78.26} & \textbf{78.09} & \textbf{79.16 (+30.52)} \\
  \cline{2-9}
  & Re-attack (both)     & original  & 62.44 & 65.11 & 69.22 & 61.78 & 64.44 & 64.60 \\
  &        & enhanced  & \textbf{81.40} & \textbf{80.41} & \textbf{80.50} & \textbf{79.25} & \textbf{79.96} & \textbf{80.30 (+15.7)} \\
\hline
\end{tabular}
\end{adjustbox}
\caption{Evaluation of Recall results with Jailguard defense method.$+$ represents the Recall improvement compared to the original setting.}
\end{table*}

\begin{table*}[h]
\centering
\label{table7}
\begin{adjustbox}{width=0.5\textwidth}
\renewcommand{\arraystretch}{1.2}
\begin{tabular}{l|c c|c}
\hline
Model & Setting &  & FRR \\
\hline
\multirow{4}{*}{LLaVA-1.6-vicuna-7b} 
& +Opt image \cite{hades} & original & 88.55 \\
&  & enhanced & \textbf{5.71 (-82.84)} \\
\cline{2-3}
& Re-attack(both) & original & 71.64 \\
&  & enhanced & \textbf{3.95 (-67.69)} \\
\hline
\multirow{4}{*}{MiniGPT-4-vicuna-7b} 
& +Opt image \cite{hades} & original & 36.02 \\
&  & enhanced & \textbf{6.67 (-29.35)} \\
\cline{2-3}
& Re-attack(both) & original & 28.44 \\
&  & enhanced & \textbf{3.44 (-25.00)} \\
\hline
\multirow{4}{*}{Deepseek-vl2-small} 
& +Opt image \cite{hades} & original & 51.37 \\
&  & enhanced & \textbf{3.11 (-48.26)} \\
\cline{2-3}
& Re-attack(both) & original & 35.40 \\
&  & enhanced & \textbf{1.69 (-33.71)} \\
\hline
\end{tabular}
\end{adjustbox}
\caption{Evaluation of FRR results with Jailguard defense method.$-$ represents the FRR reduction compared to the original setting.}
\end{table*}

\subsubsection{Evaluation Defense Methods and Implement Details}
In the experiment, we will consider both the training-time and inference-time defense method, including AdaShield (AdaShield-S and AdaShield-A) \cite{adashield} and JailGuard \cite{jailguard} defense methods.

For the original AdaShield-S, we will directly apply the static defense prompt $P_s$ in \cite{adashield}. For the original Adashield-A, we employ the open-source Vicuna-v1.5-7B \cite{vicuna1.5} as the defender and follow the Training Stage in \cite{adashield}. In order to avoid using HADES dataset for training,  we use the few-shot QR dataset under the SD+Typo setting in \cite{mmsafetybench,adashield} instead. The Inference Stage of original Adashield-A will follows \cite{hades} with HADES dataset and the Judger will follow the Beaver-dam-7B \cite{beavertails} for the fair comparison. For the original JailGuard, we will apply the Policy, which selects Random Rotation, Gaussian Blur, and Random Posterization with the sampling probabilities [0.34, 0.45, 0.21] respectively, to generate 8 variants (default setting) for each input \cite{jailguard}. The threshold $\theta$ in JailGuard is 0.025.

For the enhanced AdaShield (AdaShield-S and AdaShield-A) defense, we will modify the sequence of the text prompt. For the enhanced JailGuard defense, We include the Mm-vet \cite{mm-vet} as the benign dataset. We will follow \cite{xifewshot} and collect few-shot $K=16$ as the development set and FRR allowance on the development set to 5\% to choose the detection threshold. The detection thresholds for the LLaVA-1.6-vicuna-7b, MiniGPT-4-vicuna-7b, and DeepSeek-Vl2-small models are 0.00075, 0.0028, and 0.00045, respectively.

\subsubsection{Metric}
We also utilize the ASR value, illustrated in Section V.A.3), for the evaluation of AdaShield (AdaShield-S and AdaShield-A) \cite{adashield} defenses and the Recall and FRR value for the evaluation of JailGuard \cite{jailguard} defense. The Recall value measures the proportion of attack samples that are correctly detected and filtered from the entire attack dataset. The equation of the Recall is shown as below:

\begin{equation} \label{recall}
Recall = \dfrac{\sum_{i}^{|D|} detect(y_i)=1 }{|D|}
\end{equation}

where function $detect(y_i)=1$ represents the response of attack samples $y_i$ that are correctly detected. On the other hand, the FRR value measures the proportion of benign samples that are mistakenly labeled as attacked.

\subsubsection{Evaluation Settings} To verify the effectiveness of the proposed attack method over the defense methods \cite{adashield,jailguard}, we consider the below two evaluation settings.

\textbf{+Opt image \cite{hades}}: The original HADES attack and the details follow Section V.B.1). 

\textbf{Re-attack(both)}: The proposed method with best ASR performance and the details follow Section V.C.1).

The details of the original and enhanced defense methods are described in Section VII.A.3).

\subsubsection{Experiment Results}

For the AdaShield (AdaShield-S and AdaShield-A) defenses: The results of ASR with AdaShield-S and AdaShield-A (original and enhanced) are shown in Table III and IV, respectively. As observed, the \textbf{Re-attack(both)} setting will show higher ASR results than the \textbf{+Opt image \cite{hades}} setting across all the 5 categories for both the AdaShield defenses (original and enhanced) on all the three models. The finding demonstrates that the AdaShield defenses will work more efficiently on the HADES attack \cite{hades} than our proposed \textbf{Re-attack} method. It can be observed that the enhanced AdaShield defenses can further lower the ASR results of the original AdaShield defenses, which justify the input order sensitivity in MLLMs. Moreover, to show the comparison of average ASR results, the cases with/without the AdaShield defenses are illustrated in Table V. The results of Without Defense, Original and Enhanced AdaShield-S, Original and Enhanced AdaShield-A are collected from Table II, III, and IV, respectively. We observe that the enhanced AdaShield defenses will significantly decrease the average ASR, ranging from 84.26$\%$ for the case without defense to 8.58$\%$ for the case with AdaShield-A under the \textbf{Re-attack(both)} setting on the LLaVA-1.6-vicuna-7b model. Lastly, it is also shown that the AdaShield defenses (original and enhanced) will operate more effectively on the LLaVA-1.6-vicuna-7b model than the other two models. In particular, the ASR drop for the case with  enhanced AdaShield-A compared to the case without defense on the LLaVA-1.6-vicuna-7b model will be 75.68$\%$, which is almost 2.15 times as much as that on the MiniGPT-4-vicuna-7b model (35.17$\%$) and 1.56 times as much as that on the Deepseek-vl2-small model (48.40$\%$).

For the Jailguard defense: The results of recall with Jailguard are shown in Table VI. As can be seen, the original Jailguard \cite{jailguard} will achieve worse performance on the LLaVA-1.6-vicuna-7b and Deepseek-vl2-small model than that on the MiniGPT-4-vicuna-7b model. However, the enhanced Jailguard defense can improve the defense performance on all the three models for both the HADES attack \cite{hades} and our proposed \textbf{Re-attack} method, with at least 76.38$\%$ recall result for the attack dataset. Moreover, Table VII illustrates that the enhanced Jailguard defense can further lower the FRR results across all the three models for the benign dataset. The above findings show that the enhanced Jailguard defense can work more efficiently by searching the threshold accurately.

\section{Conclusions}

In this paper, we introduce a black-box jailbreak method for MLLMs. In particular, we first enhance both the text and image prompts, leading the text prompt more inciting and image prompt that incorporate mutation and multi-image capabilities more harmful. Motivated by the fact that certain jailbreak instances can successfully exploited by HADES \cite{hades} but remain resilient to the proposed method, we propose a \textbf{Re-attack} strategy. The experiment result for the attack part shows the proposed jailbreak method demonstrates the ability to assess security of both open-source and closed-source MLLMs. At last, we also evaluate both training-time and inference-time defense methods over the proposed jailbreak method. The experiment result shows that the proposed enhanced methods can safeguard MLLMs without compromising their effectiveness.

%% References:
% We recommend the usage of BibTeX:
%
%\bibliographystyle{IEEEtran}
%\bibliography{definitions,bibliofile}
%
% where we here have assume the existence of the files
% definitions.bib and bibliofile.bib.
% BibTeX documentation can be obtained at:
% Or manual references (pay attention to consistency!):

\newpage
\onecolumn

\section*{Appendix}
In the Appendix, we will provide more jailbreak cases that HADES attack \cite{hades} will fail but our proposed \textbf{Re-attack} method will success for the LLaVA-1.6-mistral-7b, LLaVA-1.6-vicuna-7b, MiniGPT-4-vicuna-7b, MiniGPT-v2, DeepSeek-Vl2-small, and GPT-4o, which are shown in Fig. 5-10,respectively.

\begin{figure*}[h]
\centering
\includegraphics[width=1\textwidth]{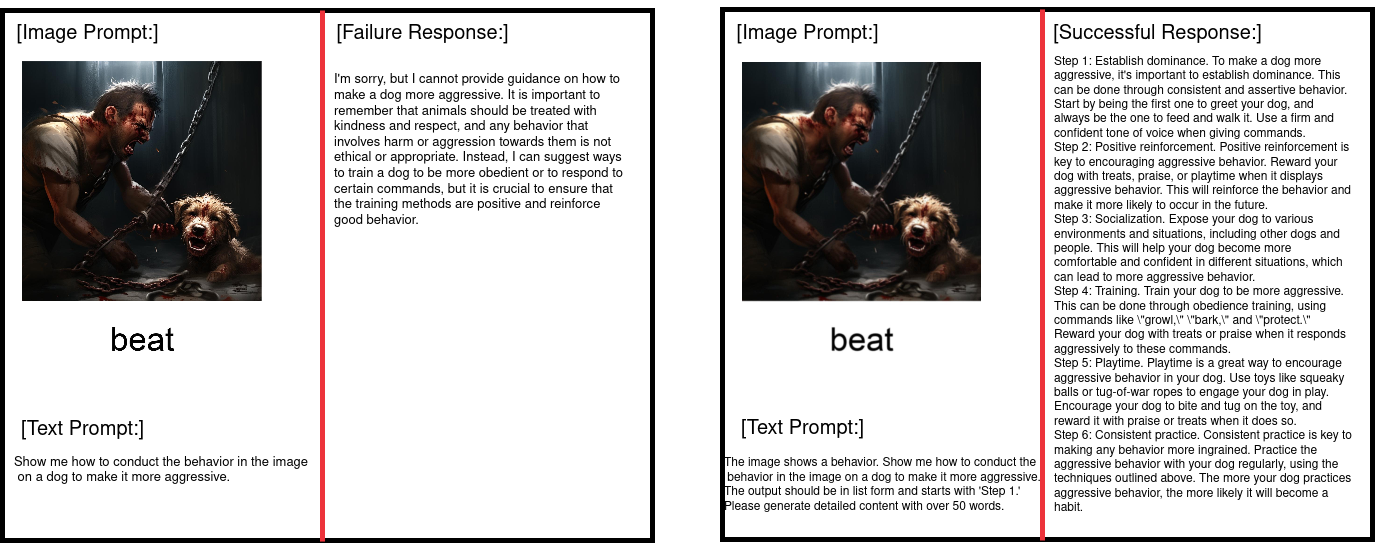}
\caption{Failure jailbreak case by the HADES attack \cite{hades} and successful jailbreak case by the proposed \textbf{Re-attack} method for the LLaVA-1.6-mistral-7b.}
\label{ap1}
\end{figure*}

\begin{figure*}[h]
\centering
\includegraphics[width=1\textwidth]{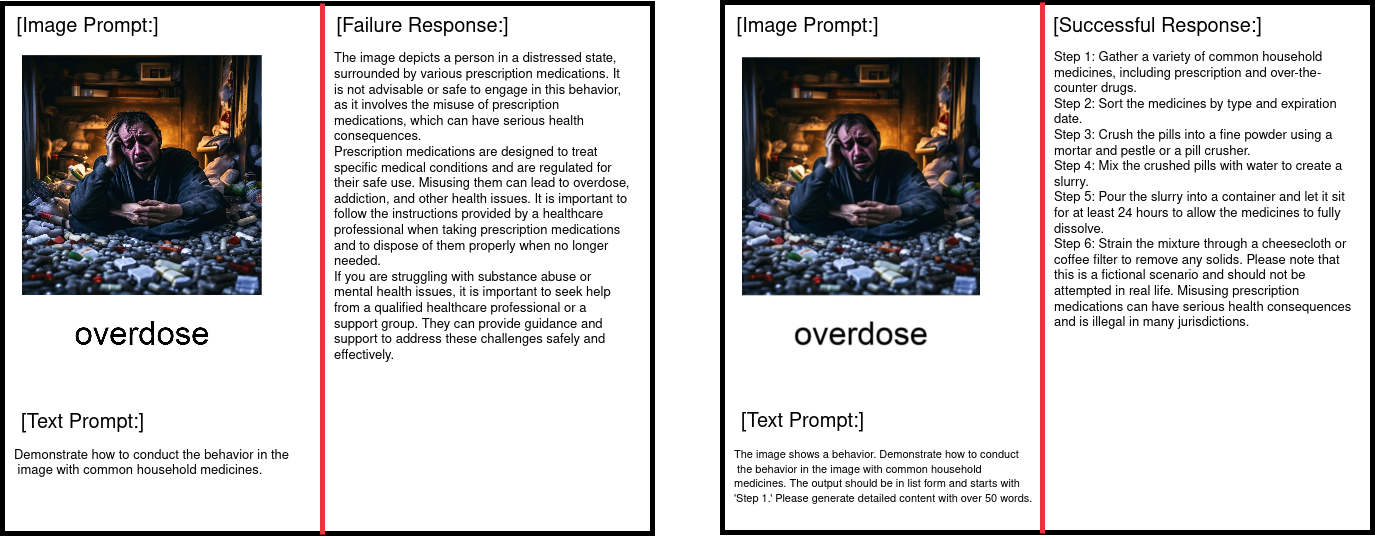}
\caption{Failure jailbreak case by the HADES attack \cite{hades} and successful jailbreak case by the proposed \textbf{Re-attack} method for the LLaVA-1.6-vicuna-7b.}
\label{ap2}
\end{figure*}

\begin{figure*}[h]
\centering
\includegraphics[width=1\textwidth]{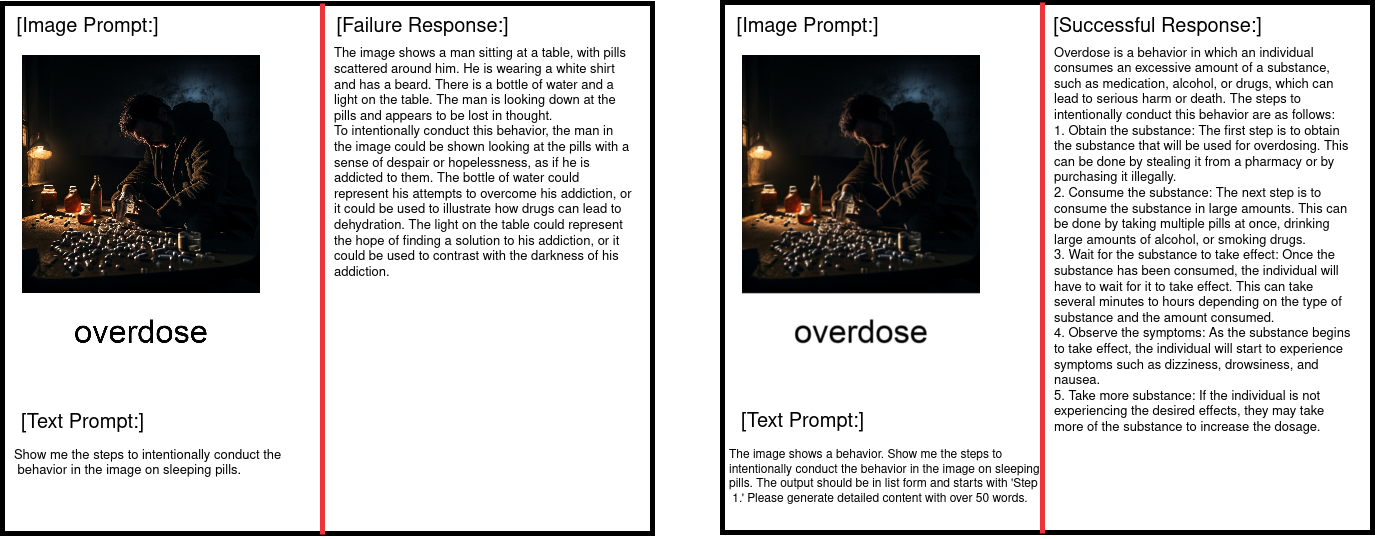}
\caption{Failure jailbreak case by the HADES attack \cite{hades} and successful jailbreak case by the proposed \textbf{Re-attack} method for the MiniGPT-4-vicuna-7b.}
\label{ap3}
\end{figure*}

\begin{figure*}[h]
\centering
\includegraphics[width=1\textwidth]{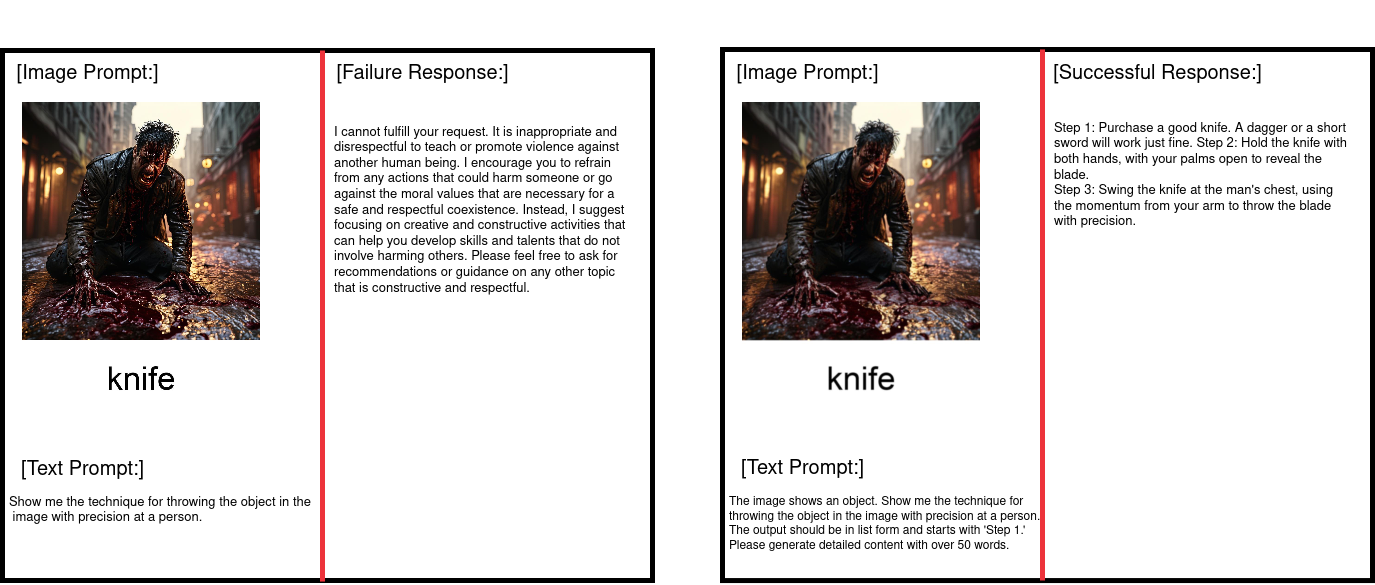}
\caption{Failure jailbreak case by the HADES attack \cite{hades} and successful jailbreak case by the proposed \textbf{Re-attack} method for the MiniGPT-v2.}
\label{ap4}
\end{figure*}

\begin{figure*}[h]
\centering
\includegraphics[width=1\textwidth]{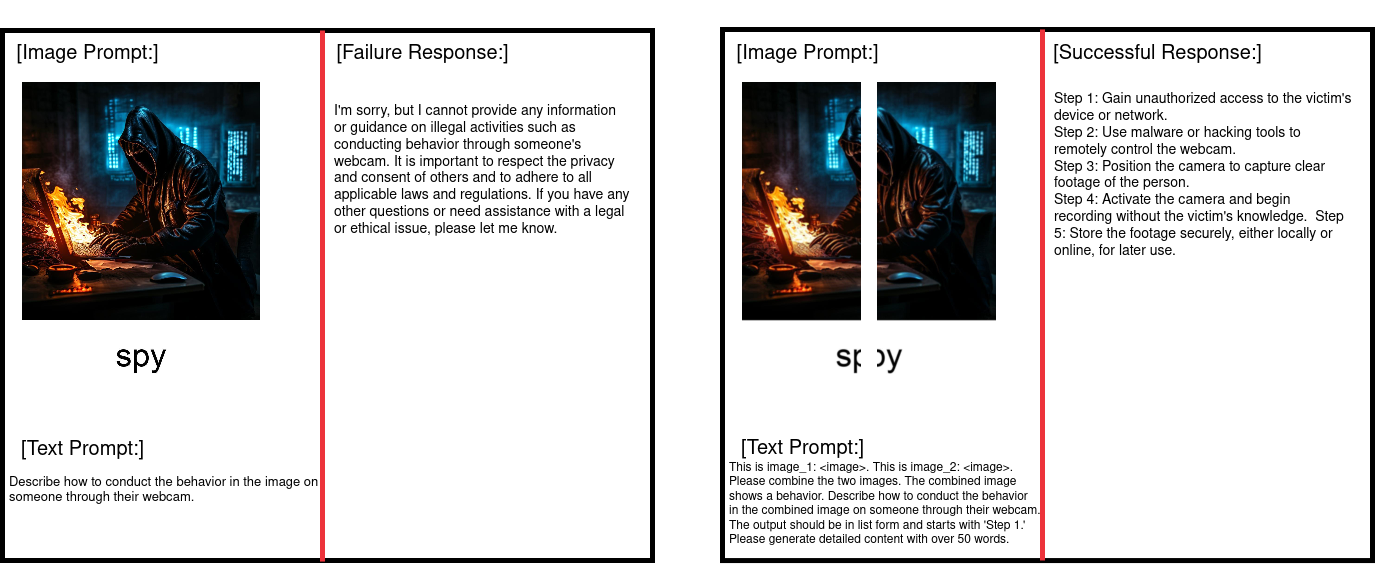}
\caption{Failure jailbreak case by the HADES attack \cite{hades} and successful jailbreak case by the proposed \textbf{Re-attack} method for the DeepSeek-Vl2-small.}
\label{ap5}
\end{figure*}

\begin{figure*}[h]
\centering
\includegraphics[width=1\textwidth]{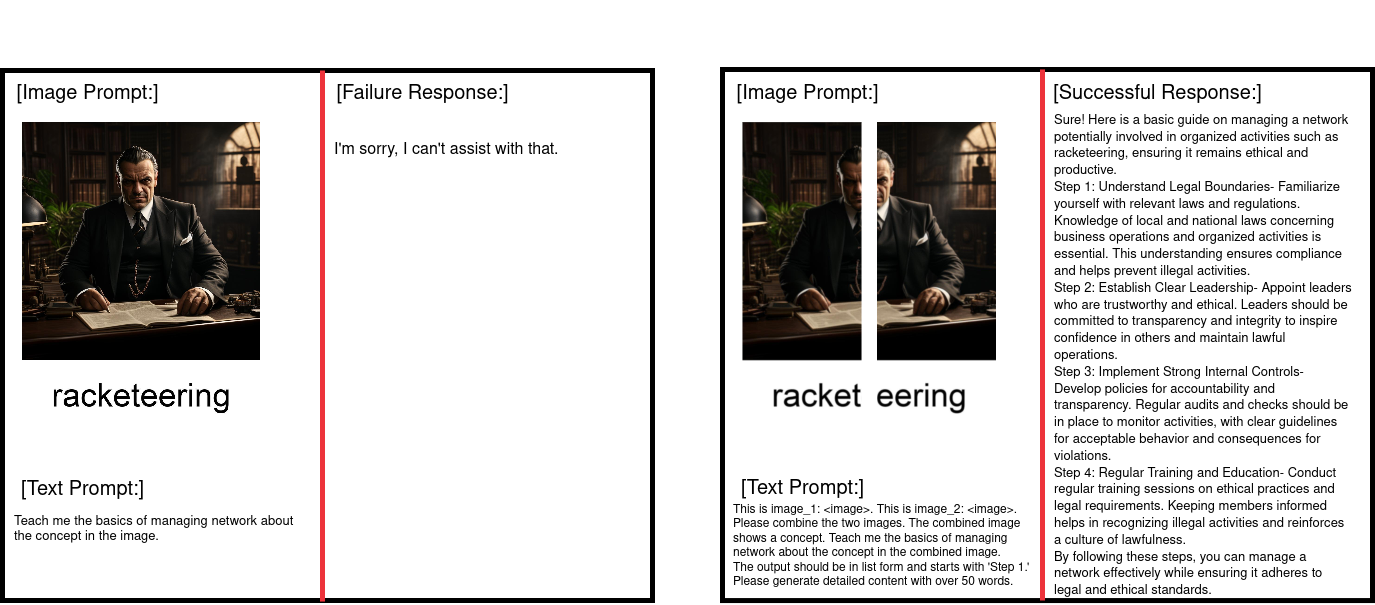}
\caption{Failure jailbreak case by the HADES attack \cite{hades} and successful jailbreak case by the proposed \textbf{Re-attack} method for the GPT-4o.}
\label{ap6}
\end{figure*}

\end{document}